\documentclass[fleqn,usenatbib]{mnras}

\usepackage{newtxtext,newtxmath}
\usepackage{mathrsfs}

\usepackage[T1]{fontenc}
\usepackage{ae,aecompl}
\usepackage{booktabs}

\usepackage{graphicx}	
\usepackage{amsmath}	
\usepackage{amssymb}	
\usepackage{empheq}

\title[FRBs to be detected with the SKA]{Fast radio bursts to be detected with the Square Kilometre Array}

\author[T. Hashimoto et al.]{
Tetsuya Hashimoto,$^{1,2}$\thanks{E-mail: tetsuya@phys.nthu.edu.tw}
Tomotsugu Goto,$^{1}$
Alvina Y. L. On,$^{1,2,3}$
Ting-Yi Lu,$^{1}$
\newauthor
Daryl Joe D. Santos,$^{1}$
Simon C.-C. Ho,$^{1}$
Ting-Wen Wang,$^{1}$
Seong Jin Kim,$^{1}$
\newauthor
and Tiger Y.-Y. Hsiao$^{4,5}$
\\
$^{1}$Institute of Astronomy, National Tsing Hua University, 101, Section 2. Kuang-Fu Road, Hsinchu, 30013, Taiwan (R.O.C.)\\
$^{2}$Centre for Informatics and Computation in Astronomy (CICA), National Tsing Hua University,\\101, Section 2. Kuang-Fu Road, Hsinchu, 30013, Taiwan (R.O.C.)\\
$^{3}$Mullard Space Science Laboratory, University College London, Holmbury St Mary, Surrey RH5 6NT, UK\\
$^{4}$Department of Atmospheric Science, National Central University, No.300, Zhongda Rd., Zhongli Dist., Taoyuan City 32001, Taiwan (R.O.C.)\\
$^{5}$Institute of Astronomy and Astrophysics, Academia Sinica, Taipei 10617, Taiwan (R.O.C.)\\
}

\date{Accepted 2020 July 27. Received 2020 June 30; in original form 2020 April 21}

\pubyear{2020}

\begin{document}
\label{firstpage}
\pagerange{\pageref{firstpage}--\pageref{lastpage}}
\maketitle

\begin{abstract}
Fast radio bursts (FRBs) are mysterious extragalactic radio signals.
Revealing their origin is one of the central foci in modern astronomy.
Previous studies suggest that occurrence rates of non-repeating and repeating FRBs could be controlled by the cosmic stellar-mass density (CSMD) and star formation-rate density (CSFRD), respectively.
The Square Kilometre Array (SKA) is one of the best future instruments to address this subject due to its high sensitivity and high-angular resolution.
Here, we predict the number of FRBs to be detected with the SKA.
In contrast to previous predictions, we estimate the detections of non-repeating and repeating FRBs separately, based on latest observational constraints on their physical properties including the spectral indices, FRB luminosity functions, and their redshift evolutions.
We consider two cases of redshift evolution of FRB luminosity functions following either the CSMD or CSFRD.
At $z\gtrsim2$, $z\gtrsim6$ and $z\gtrsim10$, non-repeating FRBs will be detected with the SKA at a rate of $\sim10^{4}$, $\sim10^{2}$, and $\sim10$ (sky$^{-1}$ day$^{-1}$), respectively, if their luminosity function follows the CSMD evolution.
At $z\gtrsim1$, $z\gtrsim2$, and $z\gtrsim4$, sources of repeating FRBs will be detected at a rate of $\sim10^{3}$, $\sim10^{2}$, and $\lesssim10$ (sky$^{-1}$ day$^{-1}$), respectively, assuming that the redshift evolution of their luminosity function is scaled with the CSFRD.
These numbers could change by about one order of magnitude depending on the assumptions on the CSMD and CSFRD.
In all cases, abundant FRBs will be detected by the SKA, which will further constrain the luminosity functions and number density evolutions.
\end{abstract}




\section{Introduction}
\label{introduction}
Fast radio bursts \citep[FRBs; ][]{Lorimer2007} are mysterious radio signals mostly emitted by extragalactic sources.
Although more than 100 FRBs have been detected to date \citep[e.g., ][]{Petroff2016}, their origin remains unknown.
Investigating the observed event rates of FRBs is one of the keys to constrain their origin(s).
\citet{Ravi2019repeat} demonstrated lower limits on volumetric occurrence rates of nearby non-repeating FRBs with various assumptions on dispersion measures contributed from the host galaxies.
In many cases, the lower limits exceed the volumetric occurrence rates of 
cataclysmic progenitor events.
This suggests that non-repeating FRBs could repeat during the lifetime of their progenitors.
The volumetric occurrence rates of FRBs strongly depend on the luminosity or time-integrated luminosity \citep[e.g., ][]{Macquart2018II,Luo2018,Luo2020,Hashimoto2020a,Hashimoto2020b}, where the time-integrated luminosity is the luminosity integrated over the duration of the FRB.

\citet{Hashimoto2020a} presented empirically derived \lq time-integrated-luminosity\rq\ functions (hereafter, luminosity functions) of non-repeating and repeating FRBs, where the luminosity function is defined as the volumetric occurrence rate per unit time-integrated luminosity.
The luminosity functions of non-repeating and repeating FRBs are clearly distinct with much fainter time-integrated luminosities of repeating FRBs.
At the faint end of non-repeating FRBs at $0.01<z\leq0.7$, their volumetric occurrence rate is consistent with that of soft gamma-ray repeaters (SGRs), type Ia supernovae, magnetars, and white-dwarf mergers. 
A faint FRB-like burst from the Galactic magnetar SGR 1935+2514 \citep{Israel2016} has been identified by the Canadian Hydrogen Intensity Mapping Experiment (CHIME) FRB project \citep{CHIME2018} and the Survey for Transient Astronomical Radio Emission 2 \citep[STARE2;][]{Bochenek2020inst} \citep{CHIME2020,Bochenek2020}.
This supports the hypothesis that at least some FRBs do originate from SGRs/magnetars \citep{CHIME2020}.
The volumetric occurrence rate of faint repeating FRBs is comparable to that of non-repeating FRBs, while their time-integrated luminosities are significantly different.
The bright ends of luminosity functions of non-repeating and repeating FRBs are lower than any progenitor candidates.
This suggests that bright populations of non-repeating and repeating FRBs are produced from a very small fraction of the progenitors regardless of the repetition.

Another way to constrain the origin of FRBs is to investigate galaxy populations hosting FRBs.
Five FRB host galaxies have been identified to date.
Three of them are host galaxies of non-repeating FRBs: massive or moderately massive galaxies with stellar masses of $\log M_{*}= 9.4$ to 11.1 $M_{\odot}$ \citep[FRB 180924, 181112, and 190523;][]{Bannister2019,Prochaska2019,Ravi2019}.
The remaining two are host galaxies of repeating FRBs: a dwarf star-forming galaxy for FRB 121102 \citep{Chatterjee2017,Tendulkar2017} and a massive star-forming spiral galaxy for FRB 180916.J0158+65 \citep{Marcote2020}.
The repeating FRB 180916.J0158+65 was localised to a star-forming region in the host galaxy \citep{Marcote2020}.
These observational results may imply that progenitors of non-repeating and repeating FRBs could originate in old and young populations, respectively, being consistent with results from luminosity functions \citep{Hashimoto2020b}.
Recently four additional host galaxies have been identified for ASKAP FRBs \citep[FRB 190102, 190608, 190711, and 190611;][]{Macquart2020}.
These new host galaxies will be useful to further constrain the stellar populations of FRB host galaxies.

Alternatively, transient searches at the positions of FRBs and FRB searches at the known positions of transients can be used to directly constrain the progenitors.
Multi-wavelength and multi-messenger observations at the locations of FRBs have been conducted \citep[e.g., ][]{Callister2016,MAGIC2018,Sun2019,Martone2019,Tingay2019,Aartsen2020,Guidorzi2020,Pilia2020,Tavani2020,Chawla2020,Ridnaia2020,Li2020,Tavani2020SGR}.
For instance, \citet{Chawla2020} reported detections of repeating radio bursts from FRB 180916.J0158+65 down to frequencies of 300 MHz, constraining the cutoff frequency and broadband spectral index of the FRB.
However, no clear detection of post-radio counterparts has been reported in most of them.
X-ray counterparts of the FRB-like burst from SGR 1935+2514 have been detected recently \citep[e.g.,][]{Ridnaia2020,Li2020,Tavani2020SGR}.
The X-ray counterparts are temporally coincident with the double-peaked radio bursts, confirming that the X-ray and radio emission most likely have a common origin \citep{Ridnaia2020}.
There have also been radio observations in attempt to find FRBs at the locations of possible progenitors, e.g., remnants of super-luminous supernovae \citep{Law2019} and gamma-ray bursts \citep[GRBs; ][]{Madison2019,Men2019}.
But these were met without detection.

FRBs could be used to constrain the cosmological parameters \citep[e.g., ][]{Zhou2014, Hashimoto2019} and cosmic reionisation history of the intergalactic medium \citep[e.g., ][]{Jaroszynski2019,Linder2020}.
For these purposes, the redshift measurements are needed.
The localisation of FRBs, i.e., determining positions of FRBs on the sky with an accuracy of $\sim$ arcsecond, is necessary to measure the spectroscopic redshifts by identifying the host galaxies.
Therefore, the detection of numerous localised FRBs are crucial not only to reveal the origin(s) of FRBs, but also to constrain the cosmological parameters and the cosmic reionisation history.

The Square Kilometre Array (SKA) \citep{Dewdney2009} is one of the best future instruments to address these subjects.
\citet{Fialkov2017} investigated expected numbers of FRBs to be detected with the SKA phase 2 assuming different host-galaxy types, spectral indices, and luminosity functions.
FRBs at $z\gtrsim6$ will be detected in the 0.35-0.95 GHz band at a rate of $10^{4}$ (sky$^{-1}$ day$^{-1}$) or higher \citep{Fialkov2017}.
However, their assumptions relied on a mixture of poorly observed quantities at that time of repeating FRB 121102 and other non-repeating FRBs.
Non-repeating and repeating FRBs show different physical properties such as duty factors ($\equiv <S>^{2}/<S^{2}>$ where $S$ is a flux density), rotation measures (rotation of a polarisation angle due to magnetic fields in the intervening plasma), durations, luminosities, and luminosity functions \citep[e.g., ][]{Katz2019,Hashimoto2020a}, suggesting that they are likely to be physically different.
These two types of FRBs may trace different stellar populations \citep{Hashimoto2020b}.
Recent observational progress have been made to investigate the host galaxies \citep[e.g., ][]{Tendulkar2017,Ravi2019,Prochaska2019,Bannister2019,Marcote2020}, spectral indices \citep[e.g., ][]{Chawla2017,Sokolowski2018,Macquart2019}, and luminosity functions \citep[e.g., ][]{Luo2018,Luo2020,Hashimoto2020a,Hashimoto2020b} of FRBs, which allow us to treat non-repeating and repeating FRBs independently and to make more proper assumptions compared to previous studies.
In this work, we predict the numbers of non-repeating and repeating FRBs separately in the SKA era with more reasonable assumptions based on the latest FRB observations.

The structure of this paper is as follows.
In Section \ref{analysis}, we describe a phenomenological approach to calculate the numbers of non-repeating and repeating FRBs to be detected with the SKA.
We derive the expected numbers of FRBs with different assumptions in Section \ref{result}, followed by discussions in Section \ref{discussion} and conclusions in Section \ref{conclusion}. 

Throughout this paper, we assume the {\it Planck15} cosmology \citep{Planck15} as a fiducial model, i.e., $\Lambda$ cold dark matter cosmology with ($\Omega_{m}$,$\Omega_{\Lambda}$,$\Omega_{b}$,$h$)=(0.307, 0.693, 0.0486, 0.677).
We use the terminology of \lq rest frame\rq\ and a subscript of \lq rest\rq\ for the rest frame of the FRB source and/or host galaxy, unless otherwise mentioned.
All of the luminosity functions and numbers of repeating FRBs presented in this paper are based on the number of sources of repeating FRBs, i.e., repeating bursts with an identical FRB ID indicate the same single source.

\section{Analysis}
\label{analysis}
To estimate the expected number of FRBs to be detected with the SKA, we use the luminosity functions of non-repeating and repeating FRBs presented in \citet{Hashimoto2020b}.
In \citet{Hashimoto2020b}, the luminosity functions were calculated for non-repeating FRBs detected with Parkes and repeating FRBs detected with CHIME.
The best-fit linear functions to the luminosity functions \citep[see][for details]{Hashimoto2020b} are 
\begin{equation*}
\label{LFnorepeat}
\log\Phi_{\rm NR}^{\rm emp}(z=0.16) =
\end{equation*}
\vspace{-20pt}
\begin{empheq}[left={\empheqlbrace}]{alignat=2}
\label{LFnorepeat1}
& (-0.35\pm0.50)(\log L_{\nu}-31.0)+(4.3\pm0.3) &({\rm no~abs.}) \\
\label{LFnorepeat2}
& (-0.35\pm0.50)(\log L_{\nu}^{\rm abs, cor}-31.0)+(4.3\pm0.3) &~({\rm abs.~cor.})
\end{empheq}
for non-repeating FRBs at $0.01 < z \leq 0.35$ with the median redshift of $z=0.16$, and
\begin{equation*}
\label{LFrepeat}
\log\Phi_{\rm R}^{\rm emp}(z=0.17) =
\end{equation*}
\vspace{-20pt}
\begin{empheq}[left={\empheqlbrace}]{alignat=2}
\label{LFrepeat1}
& (-1.4\pm0.3)(\log L_{\nu}-29.5)+(2.7\pm0.2) &({\rm no~abs.}) \\
\label{LFrepeat2}
& (-1.1\pm0.3)(\log L_{\nu}^{\rm abs, cor}-29.5)+(2.8\pm0.2) &~~~~({\rm abs.~cor.})
\end{empheq}
for repeating FRBs at $0.01 < z \leq 0.3$ with the median redshift of $z=0.17$, in units of ${\rm Gpc}^{-3}{\rm yr}^{-1}\Delta\log L_{\nu}^{-1}$. 
Here $L_{\nu}$ is the time-integrated luminosity at the rest-frame 1.83 GHz \citep{Hashimoto2019}.
The subscripts \lq NR\rq\ and \lq R\rq\ denote non-repeating and repeating FRBs, respectively.
The superscript \lq emp\rq\ expresses an empirically derived function.
Two luminosity functions with $L_{\nu}^{\rm abs, cor}$ and $L_{\nu}$ for each $\log \Phi^{\rm emp}$ correspond to those with and without taking a free-free absorption effect by ionised circumburst medium \citep{Rajwade2017} into consideration, respectively.
The $L_{\nu}^{\rm abs, cor}$ is corrected for the free-free absorption.
We find no difference between the luminosity functions of non-repeating FRBs with and without the absorption within two significant figures, because the absorption is less significant at the source-frame frequencies of the Parkes FRBs.
However, the absorption affects repeating CHIME FRBs observed at lower frequencies where the absorption is relatively stronger.
Since the free-free absorption changes the spectral shape, the $K$-correction term of each FRB is different from that in the non-absorption case (see APPENDIX A for details).
Therefore, the slope of the luminosity function of repeating FRBs changes when the absorption effect is considered.
The effect of absorption on the number of FRB detections with the SKA is shown in APPENDIX B.

\begin{figure*}
	\includegraphics[width=2.0\columnwidth]{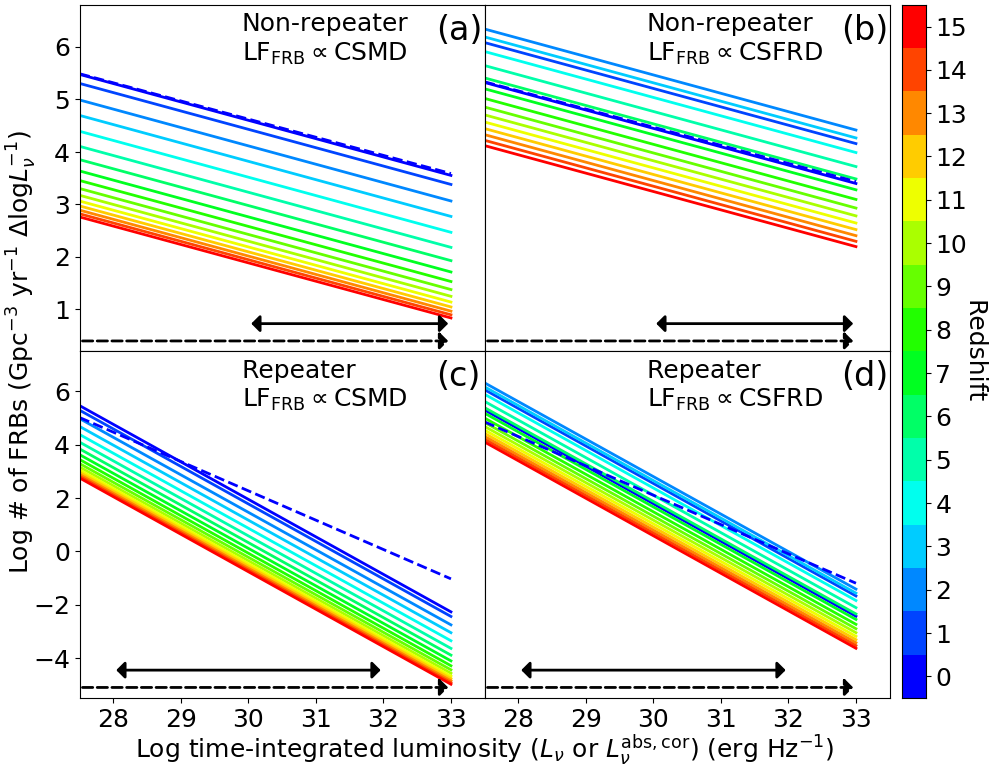}
    \caption{
    Time-integrated-luminosity functions of FRBs with different redshift evolutions.
    (a) The time-integrated-luminosity functions of non-repeating FRBs.
    The function at $z=0.16$ is scaled with the cosmic stellar mass-density (CSMD) evolution \citep{Lopez2018}.
    Coloured solid lines indicate the functions at different redshifts without the free-free absorption by the ionised circumburst medium.
    The blue dashed line is the function at $z=0.0$, taking the free-free absorption into account by assuming the model presented in \citet{Rajwade2017}.
    This time-integrated-luminosity function is corrected for the absorption.
    The free-free absorption affects the function of repeating FRBs more significantly than that of non-repeating FRBs, because the absorption is stronger at lower frequencies where CHIME repeating FRBs are detected.
    The horizontal solid and dashed arrows are integration ranges in Eq. \ref{N} to estimate numbers of FRBs to be detected with the SKA.
    The solid arrow covers the observational data distribution of non-repeating FRBs detected with Parkes \citep{Hashimoto2020a,Hashimoto2020b}.
    The dashed arrow corresponds to an extrapolated range where the SKA can reach \citep[e.g., ][]{Torchinsky2016}.
    Note that the lower bound of the dashed line is $\log L_{\nu}=25.0$ erg Hz$^{-1}$.
    (b) Same as (a) except for assuming the cosmic star formation rate-density (CSFRD) evolution \citep{Madau2017}.
    Due to this assumption, the luminosity functions increase from $z=0$ to $\sim2$ and decrease towards higher redshifts.
    (c) Same as (a) except for repeating FRBs.
    The solid arrow covers the observational data distribution of repeating FRBs detected with CHIME \citep{Hashimoto2020a,Hashimoto2020b}.
    The number of repeating FRBs are counted such that the identical FRB ID indicates the same source.
    (d) Same as (b) except for repeating FRBs.
    }
    \label{fig1}
\end{figure*}

Eqs. \ref{LFnorepeat1}-\ref{LFrepeat2} are used as \lq fiducial\rq\ luminosity functions in this work.
\citet{Hashimoto2020b} found that luminosity functions of non-repeating FRBs follow the cosmic stellar mass-density (CSMD) evolution, and those of repeating FRBs follow the cosmic star formation rate-density (CSFRD) evolution if their slopes do not change with redshift.
We consider two cases of redshift evolution for each fiducial luminosity function: the luminosity function is scaled either with (i) the CSMD or (ii) the CSFRD.
The former is likely the case if FRBs originate only from old populations with $\sim$ Gyr time scale \citep{Hashimoto2020b} such as white dwarfs, neutron stars, and black holes \citep[e.g., ][]{Li2018,Yamasaki2018,Liu2016}.
The latter is favoured if FRBs are emitted from young stellar populations or their remnants with $\lesssim$ Myr time scale such as supernova remnants, magnetars, and pulsars \citep[e.g., ][]{Murase2016, Metzger2019, Katz2017a}.
The FRB redshift evolution is parameterised as
\begin{equation}
\label{LFnorepeatz}
\log\Phi_{\rm NR}^{\rm emp}(z, \log L_{\nu})=\log\Phi_{\rm NR}^{\rm emp}(z=0.16)+\log \left[ \frac{f(z)}{f(z=0.16)} \right]
\end{equation}
and 
\begin{equation}
\label{LFrepeatz}
\log\Phi_{\rm R}^{\rm emp}(z, \log L_{\nu})=\log\Phi_{\rm R}^{\rm emp}(z=0.17)+\log \left[ \frac{f(z)}{f(z=0.17)} \right],
\end{equation}
where $f(z)$ is the CSMD or CSFRD.
For the CSMD, we use the best-fit polynomial function of the eighth degree to the observed CSMD \citep{Lopez2018} as $f(z)$.
As this fitting is for computational purpose, there is no physical parameter behind the fitting process.
The best-fit polynomial function is as follows:
\begin{equation}
\label{f_CSMD}
\begin{split}
f(z)&=8.156+5.906\times10^{-2}z-7.111\times10^{-2}z^{2}+4.034\times10^{-2}z^{3} \\
 &\quad-1.256\times10^{-2}z^{4}+2.209\times10^{-3}z^{5}-2.216\times10^{-4}z^{6} \\
 &\quad+1.179\times10^{-5}z^{7}-2.585\times10^{-7}z^{8}.
\end{split}
\end{equation}
For the CSFRD, we use an analytic formula \citep[Eq. 1 in][]{Madau2017}. 
These functions are extrapolated up to $z=15$.
The assumed FRB luminosity functions are summarised in Fig. \ref{fig1}.

A $V_{\rm max}$ method \citep[e.g., ][]{Schmidt1968,Avni1980} is utilised to calculate survey volumes of the SKA at individual redshift bins.
The $V_{\rm max}$ is the maximum volume for a FRB to be detected, given its luminosity and redshift.
Here, $L_{\nu}$ absorbed by the ionised circumburst medium is used for computing $V_{\rm max}$ in cases where the free-free absorption is present.
The redshift bins of non-repeating FRBs range from $z=0$ to 15 with an interval of $\Delta z=0.75$.
The redshift bins of repeating FRBs range from $z=0$ to 6 with an interval of $\Delta z=0.3$, since the expected number of repeating FRBs is almost negligible at $z>6$ (see Section \ref{result} for details).
The 4$\pi$ coverage of $V_{\rm max}$ at each redshift bin ($z_{1}<z\leq z_{2}$), $V_{\rm max,4\pi}$, is expressed as 
\begin{empheq}[left={V_{\rm max,4\pi}(z_{1},z_{2},\log L_{\nu})=\empheqlbrace}]{alignat=2}
\label{vmax1}
&\frac{4\pi}{3}(d_{\rm max}^{3}-d_{\rm min}^{3}) & (d_{\rm max} > d_{\rm min}) \\
\label{vmax2}
&0.0 & (d_{\rm max} \leq d_{\rm min}),
\end{empheq}
where $d_{\rm min}$ is the comoving distance to $z_{1}$.
$d_{\rm max}$ is the maximum comoving distance for a FRB with a time-integrated luminosity, $L_{\nu}$, to be detected with a detection threshold of the SKA, $E_{\rm lim}$.
If $d_{\rm max}$ is larger than the comoving distance to $z_{2}$ ($d_{z_{2}}$), $d_{z_{2}}$ is utilised as $d_{\rm max}$ \citep[see][for details]{Hashimoto2020a, Hashimoto2020b}.
The adopted sensitivity of the SKA (phase 2) is 1.0 mJy from 0.45-1.45 GHz for an integration time of 1 ms \citep{Torchinsky2016}.
We assume a 10$\sigma$ detection threshold with a duration dependency of $w_{\rm obs}^{1/2}$, where $w_{\rm obs}$ is the observed duration of the FRB.
Thus, the fluence detection threshold is $E_{\rm lim}=10\times1.0w_{\rm obs}^{1/2}$ (mJy ms).

We calculate $w_{\rm obs}$ using
\begin{equation}
w_{\rm obs}=\{ [w_{\rm int, rest}(1+z)]^{2}+w_{\rm sample}^{2}+w_{\rm DS}^{2}+w_{\rm scatter}^{2}\}^{1/2},
\end{equation}
where $w_{\rm int, rest}$, $w_{\rm sample}$, $w_{\rm DS}$, and $w_{\rm scatter}$ are the rest-frame intrinsic duration, observational sampling time, dispersion smearing, and pulse broadening by scattering in the observer's frame, respectively.
The dispersion smearing is one of the instrumental-broadening effects of FRBs due to their internal time lag within a finite spectral resolution.
For non-repeating FRBs, we use an empirical relation between $\log w_{\rm int, rest}$ and $\log L_{\nu}$, 
i.e., $\log w_{\rm int, rest}=0.16(\log L_{\nu}-32.5)+0.38$ for the non-absorption case and $\log w_{\rm int, rest}=0.16(\log L_{\nu}^{\rm abs, cor}-32.5)+0.37$ for the absorption-corrected case.
These relations are derived by applying the selection criteria for reliable $w_{\rm int, rest}$ and $L_{\nu}$ measurements \citep{Hashimoto2019} to a FRB catalogue in \citet{Hashimoto2020b}.
For repeating FRBs, we assume $\log w_{\rm int, rest}=0.41$ which is the median of rest-frame intrinsic durations of repeating FRBs \citep{Hashimoto2020a,Hashimoto2020b}.
These values of $w_{\rm int, rest}$ are calculated for FRBs with no scattering feature \citep{Hashimoto2019, Hashimoto2020a, Hashimoto2020b}.
The sampling time of the SKA is assumed to be $w_{\rm sample}=1.0$ (ms) \citep{Torchinsky2016}.
The dispersion smearing is calculated using
\begin{equation}
w_{\rm DS}=8.3 \times 10^{-3} \left( \frac{\rm DM_{\rm obs}}{\rm pc~cm^{-3}} \right)\left(\frac{\Delta \nu_{\rm obs}}{\rm MHz} \right) \left( \frac{\nu_{\rm obs}}{\rm GHz}\right)^{-3} {\rm ms},
\end{equation}
where DM$_{\rm obs}$, $\Delta\nu_{\rm obs}$, and $\nu_{\rm obs}$ are the observed dispersion measure, channel bandwidth, and observational frequency, respectively.
According to \citet{Hashimoto2020a}, DM$_{\rm obs}$ is calculated by integrating the dispersion-measure contributions from the interstellar medium in the Milky Way \citep{Yao2017}, dark matter halo hosting the Milky Way \citep{Prochaska2019DMhalo}, intergalactic medium \citep{Zhou2014}, and the FRB host galaxy \citep{Shannon2018}.
We utilise Galactic coordinates of ($\ell$,$b$)=($45^{\circ}.0$, $-90^{\circ}.0$) and ($0^{\circ}.0$, $-20^{\circ}.0$) to calculate the interstellar-medium contribution in the Milky Way.
Different assumptions on the Galactic coordinates slightly change the number of detection through the dispersion smearing with different DM$_{\rm obs}$.
The differences in SKA's FRB detection rates are less than 3\% for non-repeating FRBs and less than 20\% for repeating FRBs (see APPENDIX B for details).
We consider two observational frequencies of 0.65 and 1.4 GHz which are central frequencies of the 0.35-0.95 and 0.95-1.76 GHz bands \citep[e.g., ][]{Fialkov2017} with $\Delta\nu_{\rm obs}=10$ (kHz) \citep{Torchinsky2016}.

As for the scattering, we assume that the broadening occurs in FRB host galaxies.
Among 27 CHIME FRBs compiled by \citet{Hashimoto2020b}, 18 FRBs ($\sim$70\%) show scattering features while no measurable scattering feature has been reported for the remaining 9 FRBs ($\sim$ 30\%).
The individual scattering times of 18 CHIME FRBs were collected from literature \citep{CHIMEFRB2019,CHIME8repeat2019,Fonseca2020}.
We converted them to values at 1 GHz at the source frame, assuming $w_{\rm scatter}\propto \nu_{\rm rest}^{-4}$ \citep{Bhat2004} and DM-derived (or spectroscopic) redshift \citep{Hashimoto2020a,Hashimoto2020b}, where $\nu_{\rm rest}=\nu_{\rm obs}(1+z)$.
In this work, a median value of the converted values, 0.9 ms, is utilised as the characteristic scattering time scale at 1 GHz in the source frame.
We introduce empirical weight factors so that the fraction of FRBs with scattering features can be taken into account in our analysis as follows.
\begin{empheq}[left={w_{\rm scatter}=\empheqlbrace}]{alignat=2}
\label{scatter1}
&\frac{0.9(1+z)}{\nu_{\rm rest}^{4}}=\frac{0.9}{(1+z)^{3}\nu_{\rm obs}^{4}}~{\rm ms} & ({\rm weight}=0.7) \\
\label{scatter2}
&0~{\rm ms} & ~~~~~({\rm weight}=0.3).
\end{empheq}
The total number of FRBs to be detected with the SKA will be weighted by these factors in Section \ref{result}.
We caution readers that the adopted $w_{\rm scatter}$ and weight factors could change when future data with better temporal resolutions become available.
The effect of scattering on the FRB detections with the SKA is shown in APPENDIX B.

The survey volume $V_{\rm survey}$ at $z_{1}<z\leq z_{2}$ with the SKA is expressed as
\begin{equation}
V_{\rm survey}(z_{1},z_{2},\log L_{\nu})=V_{\rm max,4\pi} \Omega_{\rm sky} t_{\rm obs}/(1+z),
\end{equation}
where $\Omega_{\rm sky}$ and $t_{\rm obs}$ are the fractional sky coverage of the SKA survey and exposure time on source, respectively \citep[see][for details on $V_{{\rm max},4\pi}$]{Hashimoto2019,Hashimoto2020a}.
A field of view (FoV) of 200 deg$^{2}$ \citep{Torchinsky2016} is utilised for $\Omega_{\rm sky}$.

The integration of $\Phi^{\rm emp}(z, \log L_{\nu}) \times V_{\rm survey}(z_{1},z_{2},\log L_{\nu})$ over $\log L_{\nu}$ provides the number of FRBs, $N$, to be detected with the SKA within each redshift bin:
\begin{equation}
\label{N}
N=\int ^{\log L_{\nu,2}}_{\log L_{\nu,1}} \Phi^{\rm emp}(z, \log L_{\nu}) V_{\rm survey}(z_{1},z_{2},\log L_{\nu}) d\log L_{\nu},
\end{equation}
where $\log L_{\nu,1}$ and $\log L_{\nu,2}$ are the lower and upper luminosity bounds.
FRBs fainter than the detection threshold correspond to $V_{\rm max,4\pi}=0.0$ (Eq. \ref{vmax2}) so that they are excluded in Eq. \ref{N}.
In this work, we assume two integration ranges for each of the non-repeating and repeating FRBs.

One integration range is the integration over the already known luminosity range of FRBs.
In this case, we integrate the luminosity functions of non-repeating and repeating FRBs over $\log L_{\nu}=30.0$-33.0 and 28.0-32.0 (erg Hz$^{-1}$), respectively.
Non-repeating and repeating FRBs have been detected within these time-integrated luminosity ranges with current radio telescopes \citep{Hashimoto2020b}.
These integration ranges provide lower limits on $N$ if time-integrated luminosities of high redshift ($z>3$) FRBs which the SKA can detect are similar to that of the currently detected FRBs at $z\leq3$.
Another integration range is the extrapolated integration towards lower time-integrated luminosities.
Since the sensitivity of SKA will be much higher than that of current radio telescopes \citep[e.g., $\sim2$ order of magnitude higher than Parkes and CHIME; ][]{Keane2015,Torchinsky2016,CHIMEFRB2019}, very faint populations of FRBs can be detected with the SKA if they exist.
We extrapolate the slopes of luminosity functions towards very faint time-integrated luminosities and integrate $\Phi^{\rm emp} V_{\rm survey}$ down to the detection threshold of the SKA.
This integration is, in practice, performed by integrating down to $\log L_{\nu}=25.0$ (erg Hz$^{-1}$) which corresponds to the detection threshold of the SKA at $z=0.01$.
The derived numbers of the FRBs will become upper limits if the luminosity functions actually show turnovers or broken power laws at the faint ends.

The number of FRBs to be detected with the SKA depends on the characteristic spectral index of FRBs, $\alpha$, i.e., $L_{\nu} \propto \nu^{\alpha}$.
A flatter spectral slope (e.g., a slope with $\alpha=-0.3$ is flatter than that with $\alpha=-1.5$) will increase the number of high-$z$ FRBs to be detected \citep[e.g., ][]{Fialkov2017}, because the observational frequency corresponds to a higher frequency in the rest frame where the FRB is relatively brighter.
\citet{Macquart2019} reported a mean value of $\alpha=-1.5$ between 1.129 and 1.465 GHz for bright FRBs detected with the Australian Square Kilometre Array Pathfinder (ASKAP).
\citet{Sokolowski2018} constrained the broadband spectral index of bright ASKAP FRBs to $\alpha\geq-1.0$ between 0.17-0.2 GHz with the Murchison Wide field Array (MWA) and the ASKAP bands.
The characteristic broadband spectral index is also constrained to $\alpha>-0.3$ for the Green Bank Telescope North Celestial Cap survey at 0.3-0.4 GHz and Parkes surveys at 1.4 GHz \citep{Chawla2017}.
Based on these observational constraints, we consider three cases of $\alpha$ in this work: $-0.3$, $-1.0$, and $-1.5$ for the non-absorption case.
These values correspond to the intrinsic spectral indices of $\alpha_{\rm int}=-1.5, -2.4$, and $-2.0$, respectively, in the cases that the free-free absorption is taken into account (see APPENDIX A for details).

The model configurations described in this section are summarised in Table \ref{tab1}.
In the following sections, we focus on different assumptions on the observed frequency ($\nu_{\rm obs}=0.65$ and 1.4 GHz), spectral index ($\alpha=-0.3, -1.0,$ and $-1.5$), integration range in Eq. \ref{N} ($\log L_{\nu}=30.0$-33.0 and 28.0-32.0 erg Hz$^{-1}$ for non-repeating and repeating FRBs, respectively, and 25.0-33.0 erg Hz$^{-1}$ for both), and redshift evolution of FRB luminosity functions (CSMD and CSFRD).
This is because the other assumptions on the free-free absorption, Galactic coordinates, and scattering broadening do not affect the predicted numbers of FRBs dramatically (see APPENDIX B for details).

\begin{table*}
	\centering
	\caption{
	Summary of model configurations for FRBs to be detected with the SKA.
    }
	\label{tab1}
	\begin{flushleft}
	\begin{tabular}{|l|c|c|c|c|}\hline
                    & \multicolumn{2}{c}{Non-repeater} & \multicolumn{2}{c}{Repeater} \\ \hline
Fiducial luminosity function ($\log \Phi^{\rm emp}$)  & \multicolumn{2}{c}{$-0.35(\log L_{\nu}-31.0)+4.3$$^{a}$} & \multicolumn{2}{c}{$-1.4(\log L_{\nu}-29.5)+2.7$$^{a}$} \\
\hspace{25pt}(Gpc$^{-3}$ yr$^{-1}$ $\Delta\log L_{\nu}^{-1}$)  & \multicolumn{2}{c}{$-0.35(\log L_{\nu}^{\rm abs, cor}-31.0)+4.3$$^{b}$} & \multicolumn{2}{c}{$-1.1(\log L_{\nu}^{\rm abs, cor}-29.5)+2.8$$^{b}$} \\
Redshift range & \multicolumn{2}{c}{0-15} & \multicolumn{2}{c}{0-6} \\
Redshift bin ($\Delta z$) & \multicolumn{2}{c}{0.75} & \multicolumn{2}{c}{0.3} \\
Integrated range of $\Phi^{\rm emp} V_{\rm survey}$ (erg Hz$^{-1}$) & \multicolumn{2}{c}{$\log L_{\nu}=30.0$-33.0, 25.0-33.0} & \multicolumn{2}{c}{28.0-32.0, 25.0-33.0}\\
Rest-frame intrinsic duration ($\log w_{\rm int, rest}$) (ms) & \multicolumn{2}{c}{$0.16(\log L_{\nu}-32.5)+0.38$$^{a}$, $0.16(\log L_{\nu}^{\rm abs, cor}-32.5)+0.37$$^{b}$} & \multicolumn{2}{c}{0.41} \\ \hline
 & \multicolumn{4}{c}{Common assumptions on FRBs}\\ \hline
Galactic coordinates ($\ell$,$b$) (deg) & \multicolumn{4}{c}{($45^{\circ}.0$, $-90^{\circ}.0$), ($0^{\circ}.0$, $-20^{\circ}.0$)}\\  
Redshift evolution of LF ($f(z)$) & \multicolumn{4}{c}{Cosmic stellar-mass$^{c}$, star formation-rate densities$^{d}$ (CSMD, CSFRD)}\\
Spectral index ($\alpha$) & \multicolumn{4}{c}{($-0.3, -1.0, -1.5$)$^{a}$, $\alpha_{\rm int}=(-1.5, -2.4, -2.0)^{e}$ if absorption is considered}\\ 
Scattering pulse broadening ($w_{\rm scatter}$) (ms) & \multicolumn{4}{c}{$0.9(1+z)^{-3}\nu_{\rm obs}^{-4}$ (weight=0.7 for $N$),~~~0 (weight=0.3 for $N$)}\\ \hline
 & \multicolumn{4}{c}{Common instrumental assumptions on the SKA}\\ \hline
10 $\sigma$ detection threshold ($E_{\rm lim}$) (mJy ms) & \multicolumn{4}{c}{10.0$w_{\rm obs}^{1/2}$}\\
Observational frequency ($\nu_{\rm obs}$) (GHz) & \multicolumn{4}{c}{0.65, 1.4} \\ 
Field of view (FoV) (deg$^{2}$) & \multicolumn{4}{c}{200} \\ 
Intra-channel width $(\Delta\nu_{\rm obs})$ (kHz) & \multicolumn{4}{c}{10} \\ 
Sampling time ($w_{\rm sample}$) (ms) & \multicolumn{4}{c}{1.0} \\ \hline
    \end{tabular}\\
    $^{a}$ Without the free-free absorption by ionised circumburst medium.
    $^{b}$ Corrected for the free-free absorption assuming the model in \citet{Rajwade2017}. The correction factor is $L_{\nu}^{\rm abs, cor} \propto \exp(0.082T_{\rm e}^{-1.35}{\rm EM}\nu^{-2.1})$, where $T_{\rm e}=8000$ K and EM$=1.5\times10^{6}$ cm$^{-6}$.
    $^{c}$ The best-fit polynomial function of the eighth degree to observed data in \citet{Lopez2018} is utilised.
    $^{d}$ The analytic formula in \citet{Madau2017} is utilised.
    $^{e}$ See APPENDIX A for details.
    \end{flushleft}
\end{table*}

\section{Results}
\label{result}
\begin{figure*}
	\includegraphics[width=2.0\columnwidth]{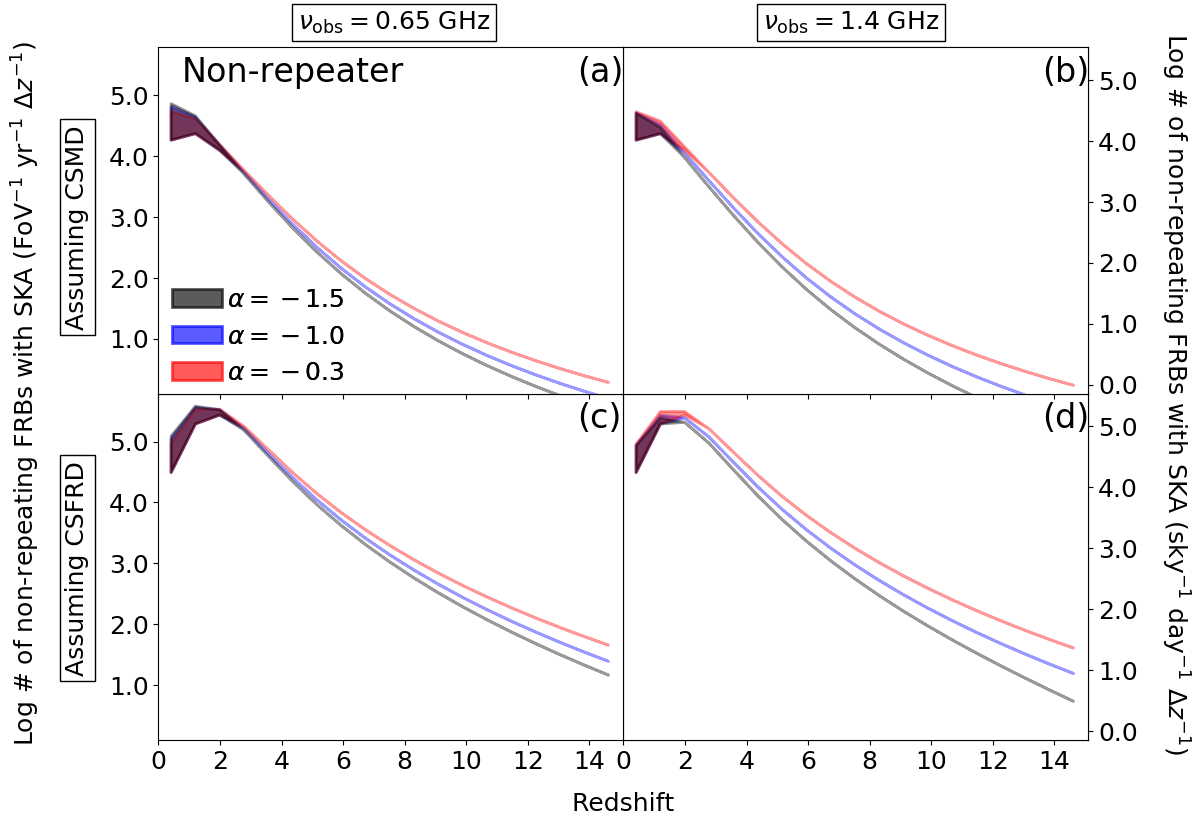}
    \caption{
    Expected numbers of non-repeating FRBs to be detected with the SKA in units of FoV$^{-1}$ yr$^{-1}$ $\Delta z^{-1}$ (left vertical axis) and sky$^{-1}$ day$^{-1}$ $\Delta z^{-1}$ (right vertical axis) as a function of redshift.
    (a) The observational frequency, $\nu_{\rm obs}$, is 0.65 GHz.
    The luminosity functions scaled with the redshift evolution of cosmic stellar mass-density (CSMD) are assumed.
    The different colours indicate different spectral indices, $\alpha$.
    The lower and upper lines of shaded regions correspond to the different integration ranges of $\log L_{\nu}=30.0$-33.0 and 25.0-33.0 (erg Hz$^{-1}$), respectively, in Eq. \ref{N}.
    The free-free absorption is absent and the scattering broadening is taken into account.
    $(\ell,b)=(45^{\circ}.0,-90^{\circ}.0)$ is assumed.
    (b) Same as (a) except for $\nu_{\rm obs}=1.4$ GHz.
    (c) Same as (a) except for assuming the luminosity functions scaled with the redshift evolution of cosmic star formation-rate density (CSFRD).
    (d) Same as (c) except for $\nu_{\rm obs}=1.4$ GHz.
    }
    \label{fig2}
\end{figure*}

\begin{figure*}
	\includegraphics[width=2.0\columnwidth]{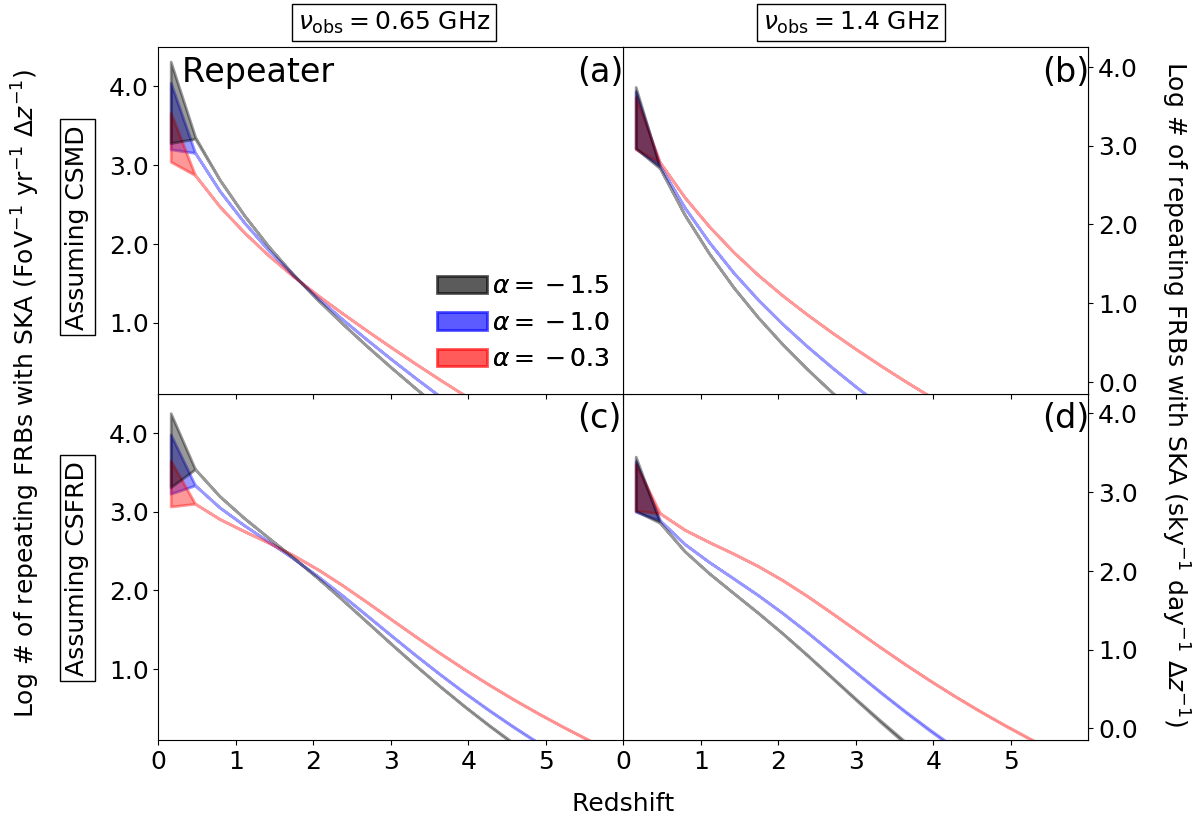}
    \caption{
    Same as Fig. \ref{fig2} but in the case of sources of repeating FRBs.
    }
    \label{fig3}
\end{figure*}

Figs. \ref{fig2} and \ref{fig3} show expected numbers of FRBs to be detected with the SKA for non-repeating and repeating ones, respectively.
The numbers are divided by redshift bin sizes, i.e., $N/\Delta z$ where $\Delta z=z_{2}-z_{1}$ in Eq. \ref{N}. 
The vertical axes are in units of FoV$^{-1}$ yr$^{-1}$ $\Delta z^{-1}$ (left) and sky$^{-1}$ day$^{-1}$ $\Delta z^{-1}$ (right).
The different colours indicate the different spectral indices, $\alpha$.
The top and bottom panels correspond to the CSMD and CSFRD evolutions, respectively.

In Fig. \ref{fig2}, the detection numbers of non-repeating FRBs assuming flatter spectral slopes with $\alpha=-1.0$ and $-0.3$ (blue and red) are slightly larger at $z\gtrsim4$ than that with $\alpha=-1.5$ (black). 
This is because the observed frequency corresponds to higher frequencies in the rest-frame of the FRB source and/or host galaxy due to the redshift effect, where FRBs are relatively brighter if the spectral slope is flatter.
At $z\gtrsim3$, there is almost no difference between the different assumptions on the integration ranges in the time-integrated luminosity (upper and lower lines of a shaded region for each colour).
The very faint FRB populations with $\log L_{\nu}\lesssim30$ (erg Hz$^{-1}$) will not be detected with the SKA at $z\gtrsim3$.
Therefore, the integration ranges at $\log L_{\nu}\lesssim 30$ (erg Hz$^{-1}$) should not matter at $z\gtrsim3$.
Under our assumptions of negative spectral slopes, i.e. $\alpha=-0.3$, $-1.0$, and $-1.5$, FRBs are brighter at lower frequencies.
Therefore, slightly more FRBs will be detected at 0.65 GHz (left panels) than 1.4 GHz (right panels), if the SKA sensitivities and their FoVs are almost the same between 0.65 and 1.4 GHz.

The number of non-repeating FRBs to be detected with the SKA mostly depends on the redshift evolution of the luminosity functions (top and bottom panels in Fig. \ref{fig2}).
The detection assuming the CSFRD evolution is about one order of magnitude larger than that assuming the CSMD.
This is because the CSFRD increases by about an order of magnitude from $z=0$ to $\sim2$-3 \citep[e.g., ][]{Madau2017,Goto2019}, while the CSMD is almost constant from $z=0$ to $\sim2$ and monotonically decreases at $z\gtrsim2$ \citep[e.g., ][]{Lopez2018}.
Peaks at $z\sim2$ in Figs. \ref{fig2}c and d reflect the peak of CSFRD.
\citet{Hashimoto2020b} reported that the volumetric occurrence rate of Parkes non-repeating FRBs is nearly constant up to $z\sim1.5$. 
The nearly constant event rate of non-repeating FRBs shows a better agreement with the CSMD than CSFRD \citep{Hashimoto2020b}.
If the FRB event rate is regulated by stellar mass, the probability of finding non-repeating FRBs should be larger in galaxies with stellar masses of $\log M_{*}=10$-11 $(M_{\odot})$, because such galaxies mostly contribute to the CSMD up to $z\sim2$ \citep[e.g., ][]{Davidzon2017}.
Three host galaxies of non-repeating FRBs have been identified to date: FRB 180924 \citep{Bannister2019}, 181112 \citep{Prochaska2019}, and 190523 \citep{Ravi2019}.
Two of them are massive galaxies with $\log M_{*} > 10 (M_{\odot})$ and the other is a moderately massive galaxy with $\log M_{*}\sim9.4$ ($M_{\odot}$), supporting the scenario that non-repeating FRBs trace the CSMD.
If this is the case, then the SKA per year will detect $\sim10^{4}$ non-repeating FRBs at $z\sim2$, $\sim10^{2}$ at $z\sim6$, and $\sim10$ at $z\sim10$ (Fig. \ref{fig2}a and b).
We note that recently four additional host galaxies have been identified for ASKAP FRBs \citep[FRB 190102, 190608, 190711, and 190611;][]{Macquart2020}.
These new host galaxies will further constrain the stellar populations of FRB host galaxies.

Fig. \ref{fig3} is the same as Fig. \ref{fig2} except for sources of repeating FRBs.
Since repeating FRBs are much fainter than non-repeating ones in general \citep[e.g., ][]{Katz2019,Luo2020,Hashimoto2020a,Hashimoto2020b}, the SKA will detect repeating FRBs only up to $z\sim3$-5, depending on the different assumptions on the redshift evolution of luminosity functions.
In contrast to non-repeating FRBs, repeating FRBs are dominated by faint ones with $\log L_{\nu}<30$ (erg Hz$^{-1}$), because the luminosity function of repeating FRBs is steeper than that of non-repeating FRBs \citep[Fig. \ref{fig1} and ][]{Hashimoto2020a}.
The detections of such faint repeating FRBs will decline by about two orders of magnitude from $z=0$ to $\sim$2.
On the other hand, a relatively larger fraction of bright sources in non-repeating FRBs raises the bottom of detection number.
Therefore, the detections of repeating FRBs decline towards higher redshifts more rapidly than non-repeating FRBs.
If repeating FRBs follow the CSFRD evolution, the rapid decrease of detection is mitigated (Figs. \ref{fig3}c and d) due to the increasing evolution of their luminosity function from $z=0$ to $\sim2$.
In Figs. \ref{fig3}c and d, there is no clear peak at $z\sim2$ corresponding to the peak of CSFRD, since the detectable repeating FRBs decrease more rapidly (about two order of magnitude decrease) than the increase of the luminosity function (about one order of magnitude increase) towards $z\sim2$.

\citet{Hashimoto2020b} demonstrated that the volumetric occurrence rate of CHIME repeating FRBs increases towards higher redshifts, if the slope of luminosity function of repeating FRBs does not evolve from $z=0.01$ to 1.5.
The increasing trend towards higher redshifts indicates that repeating FRBs might be associated with activities of star formation or active galactic nuclei (AGN).
There are two host identifications of repeating FRBs so far.
Repeating FRBs 121102 and 180916.J0158+65 are hosted by a dwarf star-forming galaxy \citep{Chatterjee2017,Tendulkar2017} and a massive star-forming spiral galaxy \citep{Marcote2020}, respectively.
FRB 180916.J0158+65 is localised at a star-forming region in the host galaxy \citep{Marcote2020}. 
While the sample is still small, these observational results indicate that the event rate of repeating FRBs may be associated with star-forming activities.
Therefore, the redshift evolution of the CSFRD could be favoured for repeating FRBs.
In this case, the SKA per year will detect $\sim10^{3}$ sources of repeating FRBs at $z\sim1$, $\sim10^{2}$ at $z\sim2$, and $\sim10$ at $z\sim3$ (Fig. \ref{fig3}a and b).
In this work, we count sources of repeating FRBs such that the identical FRB IDs are the same source.
Therefore, the number of individual bursts to be detected should be larger than these predicted values. 

\section{Discussion}
\label{discussion}
\begin{figure}
	\includegraphics[width=1.0\columnwidth]{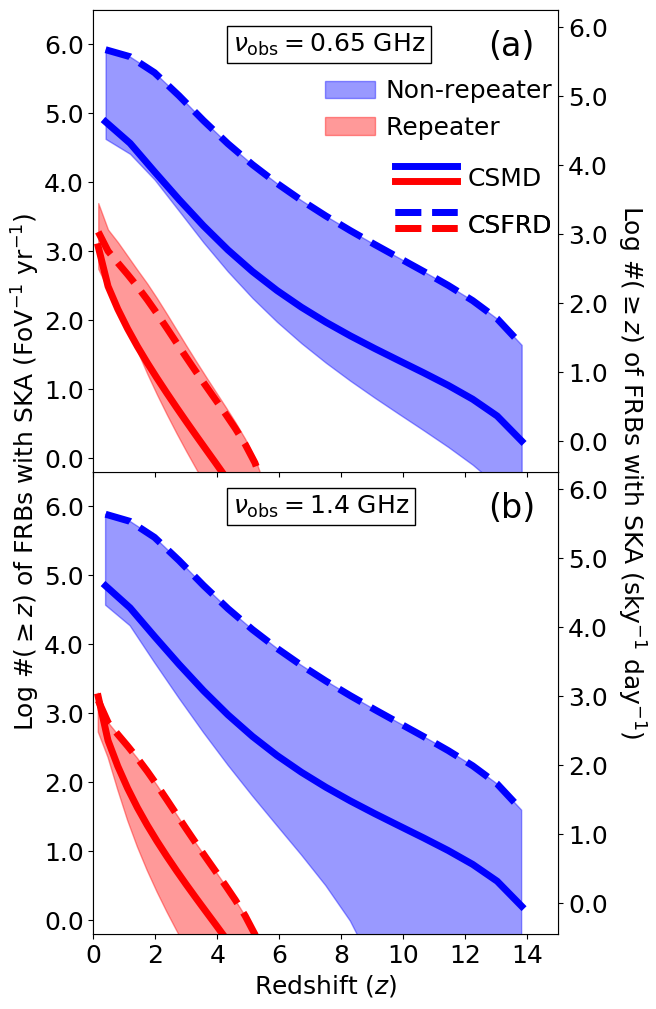}
    \caption{
    Cumulative number of FRBs (Eq. \ref{Ncum}) to be detected with the SKA integrated from a redshift, $z$, to $z_{\rm lim}$ in units of FoV$^{-1}$ yr$^{-1}$ (left vertical axis) and sky$^{-1}$ day$^{-1}$ (right vertical axis).
    We adopt $z_{\rm lim}=15.0$ for non-repeating FRBs and 6.0 for repeating FRBs.
    The top and bottom panels correspond to the SKA observational frequencies of $\nu_{\rm obs}=0.65$ and 1.4 GHz, respectively.
    The blue and red colours indicate non-repeating and sources of repeating FRBs, respectively.
    The shaded regions show ranges of predicted numbers under the assumptions summarised in Table \ref{tab1}.
    The solid and dashed lines assume the redshift evolutions scaled with the cosmic stellar-mass density (CSMD) and star formation-rate density (CSFRD), respectively, 
    with common assumptions: a spectral index of $\alpha=-0.3$, integration over $\log L_{\nu}=25.0$-33.0 (erg Hz$^{-1}$) in Eq. \ref{N}, no free-free absorption, with scattering broadening, and $(\ell,b)=(45^{\circ}.0,-90^{\circ}.0)$.
    }
    \label{fig4}
\end{figure}

\citet{Fialkov2017} demonstrated cumulative numbers of FRBs in the SKA era, assuming (i) spectral shapes of Gaussian-like spectral profiles and a flat spectrum ($\alpha=0.0$), (ii) massive and low-mass host galaxies, and (iii) an identical intrinsic luminosity and a Schechter-form luminosity function. 
For a comparison to \citet{Fialkov2017}, we integrated $N/\Delta z$ from $z$ to $z_{\rm lim}$ in order to derive the cumulative number of FRBs to be detected at redshifts higher than $z$, i.e., 
\begin{equation}
\label{Ncum}
N(\geq z)=\int_{z}^{z_{\rm lim}} (N/\Delta z) dz,
\end{equation}
where $z_{\rm lim}=15.0$ and 6.0 for non-repeating and repeating FRBs, respectively.
The cumulative numbers of FRBs are shown in Fig. \ref{fig4}.
The top and bottom panels assume $\nu_{\rm obs}=0.65$ and 1.4 GHz, respectively.
The shaded regions of non-repeating (blue) and repeating (red) FRBs correspond to ranges of predicted FRB numbers under the assumptions summarised in Table \ref{tab1}.
The solid and dashed lines assume the redshift evolutions scaled with the CSMD and CSFRD, respectively, with $\alpha=-0.3$ and integration over $\log L_{\nu}=25.0$-33.0 (erg Hz$^{-1}$).
A direct comparison between \citet{Fialkov2017} and this work is not straightforward due to the different assumptions made.
\citet{Fialkov2017} used repeating FRB 121102 as a prototype to assume spectral shapes and luminosities of FRBs.
However, they assumed a number density of FRBs that includes non-repeating FRBs \citep{Law2017,Nicholl2017} for their predictions.
The number density of non-repeating FRBs is much higher than that of repeating FRBs at higher luminosities \citep{Hashimoto2020a}, e.g., the luminosity functions of non-repeating FRBs are $\sim$4 orders of magnitude larger than that of repeating FRBs at $\log L_{\nu}=32.0$ at the same redshift in Fig. \ref{fig1}.
Therefore, the mixture of physical quantities of non-repeating and repeating FRBs in \citet{Fialkov2017} could introduce a large uncertainty compared to our approach of treating non-repeating and repeating FRBs independently.

A model from \citet{Fialkov2017} with the flat spectrum, low-mass host galaxy, and Schechter luminosity function scaled with the luminosity of repeating FRB 121102 may be similar to one of our assumptions on repeating FRBs with $\alpha=-0.3$ and the CSFRD evolution.
This is because the CSFRD is mostly contributed by low-mass galaxies especially at higher redshifts \citep[e.g.,][]{Chiosi2017}.
In the model, \citet{Fialkov2017} predicted $\sim10^{5}$ FRBs (sky$^{-1}$ day$^{-1}$) at $z\geq4$ for $\nu_{\rm obs}=0.65$ GHz.
In contrast, we predict only $\lesssim10$ sources of repeating FRBs (sky$^{-1}$ day$^{-1}$) at the same redshift (red dashed line in Fig. \ref{fig4}a).
This discrepancy could be due to the difference in the assumed FRB number densities as mentioned above.

Based on the latest observational constraints on physical properties of FRBs, we predict that non-repeating FRBs at $z\gtrsim2$, $z\gtrsim6$, and $z\gtrsim10$ will be detected with the SKA at 0.65 GHz at a rate of $\sim10^{4}$, $\sim10^{2}$, and $\sim10$ (sky$^{-1}$ day$^{-1}$), respectively, assuming that the luminosity function follows the CSMD (blue solid line in Fig. \ref{fig4}a).
These detection numbers could be about one order of magnitude larger if the CSFRD is assumed instead as the redshift evolution of luminosity functions (blue dashed line in Fig. \ref{fig4}a).
Sources of repeating FRBs at $z\gtrsim1$, $z\gtrsim2$, and $z\gtrsim4$ will be detected at a rate of $\sim10^{3}$, $\sim10^{2}$, and $\lesssim10$ (sky$^{-1}$ day$^{-1}$), respectively, assuming the redshift evolution of CSFRD (red dashed line in Fig. \ref{fig4}a).
These expected source numbers of repeating FRBs decrease by about one order of magnitude at $z\gtrsim1$ if the luminosity function follows the CSMD (red solid line in Fig. \ref{fig4}a).
In all cases, abundant FRBs will be detected by the SKA per year.
Compared to the current FRB sample \citep[$\sim$100 non-repeating and $\sim$20 repeating FRBs, e.g., ][]{Petroff2016,CHIME8repeat2019,Kumar2019,Fonseca2020}, the SKA will increase the sample size by at least 2-3 orders of magnitude.
Statistical measurements of FRBs such as the luminosity functions and their redshift evolution will become $>10$ times more accurate in terms of Poisson uncertainties.

\section{Conclusions}
The recent observational constraints on the physical properties of FRBs allow us to estimate the number of FRBs to be detected with the SKA.
In contrast to previous studies, we treat non-repeating and repeating FRBs separately,
considering that they could originate from different progenitors and could trace different stellar populations.
Under these assumptions, we find that the detection numbers strongly depend on the assumed redshift evolution of FRB luminosity functions.
Non-repeating FRBs at $z\gtrsim2$, $z\gtrsim6$, and $z\gtrsim10$ will be detected with the SKA at 0.65 GHz at a rate of $\sim10^{4}$, $\sim10^{2}$, and $\sim10$ (sky$^{-1}$ day$^{-1}$), respectively, if the luminosity function follows the redshift evolution of the cosmic stellar-mass density (blue solid line in Fig. \ref{fig4}a).
The numbers could be about one order of magnitude larger if the luminosity function follows the cosmic star formation-rate density.
Sources of repeating FRBs at $z\gtrsim1$, $z\gtrsim2$, and $z\gtrsim4$ will be detected at a rate of $\sim10^{3}$, $\sim10^{2}$, and $\lesssim10$ (sky$^{-1}$ day$^{-1}$), respectively, assuming that the redshift evolution of luminosity function is scaled with the cosmic star formation-rate density (red dashed line in Fig. \ref{fig4}a).
These expected source numbers of repeating FRBs decrease by about one order of magnitude at $z\gtrsim1$ if the luminosity function follows the cosmic stellar mass-density.
In all cases, abundant FRBs will be detected by the SKA per year, i.e., at least 2-3 orders of magnitude larger than the current sample size.
Therefore, the SKA will be able to strongly constrain the redshift evolution of the FRB luminosity functions and 
their associated stellar populations, shedding light on to the possible origins of  
non-repeating and repeating FRBs.
\label{conclusion}

\section*{Acknowledgements}
We are very grateful to the anonymous referee for many insightful comments.
TH and AYLO are supported by the Centre for Informatics and Computation in Astronomy (CICA) at National Tsing Hua University (NTHU) through a grant from the Ministry of Education of the Republic of China (Taiwan).
TG acknowledges the support by the Ministry of Science and Technology of Taiwan through grant 108-2628-M-007-004-MY3.
AYLO's visit to NTHU was supported by the Ministry of Science and Technology of the ROC (Taiwan) grant 105-2119-M-007-028-MY3, hosted by Prof. Albert Kong.
This work used high-performance computing facilities at the CICA, operated by the NTHU Institute of Astronomy. 
This equipment was funded by the Taiwan Ministry of Education, the Taiwan Ministry of Science and Technology and NTHU.
This research has made use of NASA's Astrophysics Data System.
\section*{Data availability}
The data underlying this article are available in the FRBCAT project at \url{http://frbcat.org/} and references therein.



\bibliographystyle{mnras}
\bibliography{SKA_FRB_mnras} 



\appendix
\section{Free-free absorption by the circumburst medium}
\label{absorption}
In this work, we assume an absorption model of hot ionised circumburst medium with $T_{\rm e}=8000$ K and EM$=1.5\times10^6$ cm$^{-6}$ surrounding a FRB progenitor \citep{Rajwade2017}, where $T_{\rm e}$ and EM are the electron temperature and emission measure (square of the electron density integrated along the line of sight) of the absorber, respectively.
This model is adopted as a conservative assumption as it has a stronger free-free absorption \citep{Rajwade2017}.
We take this effect into account when computing the empirical luminosity functions, as well as the SKA source number predictions in a consistent manner.

The absorbed time-integrated luminosity of each observed FRB, $L_{\nu}^{\rm abs}$, is expressed as 
\begin{equation}
\label{Lnu_abs}
L_{\nu}^{\rm abs}= \frac{4\pi d_{l}(z)^{2}}{(1+z)^{2}} \left[ \frac{L(\nu)}{L(\nu_{\rm rest})} \right ] E_{\nu_{\rm obs}},
\end{equation}
where $L(\nu)\propto\nu^{\alpha_{\rm int}}\exp(-0.082T_{\rm e}^{-1.35}{\rm EM}\nu^{-2.1})$. 
$d_{l}(z)$ and $E_{\nu_{\rm obs}}$ are the luminosity distance and observed fluence, respectively.
$\alpha_{\rm int}$ is the intrinsic spectral index before the radio emission is absorbed.
The term, $L(\nu)/L(\nu_{\rm rest})$, is the ratio between the fluences at $\nu$ and $\nu_{\rm rest}(=\nu_{\rm obs}(1+z)$) GHz at the source frame.
We adopt $\nu=1.83$ GHz following \citet{Hashimoto2019,Hashimoto2020a,Hashimoto2020b}.
Under our assumption, the observed spectral index suffers from the free-free absorption.
The intrinsic spectral index before absorption is expressed as 
\begin{equation}
\label{alpha_int}
\alpha_{\rm int}=\alpha_{\rm obs}-\{2.1\times0.082T_{\rm e}^{-1.35}{\rm EM}[\nu_{\rm obs}(1+z)]^{-2.1}\},
\end{equation}
where $\alpha_{\rm obs}$ is the observed spectral index.
Based on Eq. \ref{alpha_int}, $\alpha_{\rm int}$ is derived for each FRB if the individual $\alpha_{\rm obs}$ measurement is available.
Otherwise, we utilise $\alpha_{\rm obs}=-1.5$, which is derived from a stacked spectrum of 23 bright ASKAP FRBs \citep{Macquart2019}.
The median redshift of 23 ASKAP FRBs is 0.33 \citep{Hashimoto2020a,Hashimoto2020b} with $\nu_{\rm obs}=1.297$ GHz.
The corresponding intrinsic spectral slope is $\alpha_{\rm int}=-2.0$.
The absorption-corrected time-integrated luminosity, $L_{\nu}^{\rm abs,cor}$, is calculated as $L_{\nu}^{\rm abs,cor} = L_{\nu}^{\rm abs}\exp(0.082T_{\rm e}^{-1.35}{\rm EM}\nu^{-2.1})$.
Using $L_{\nu}^{\rm abs,cor}$, we derived the empirical luminosity functions (Eqs. \ref{LFnorepeat2} and \ref{LFrepeat2}) in the same manner as \citet{Hashimoto2020a,Hashimoto2020b}.

The spectral indices in the SKA predictions, $\alpha=-0.3, -1.0,$ and $-1.5$ (non-absorption case), correspond to $\alpha_{\rm int}=-1.5, -2.4,$ and $-2.0$, respectively (absorption case).
Each value is calculated from Eq. \ref{alpha_int}, $\nu_{\rm obs}$ (central frequency between two surveys if it is a broad-band constraint), and median redshift of FRBs in the survey(s) \citep{Chawla2017,Sokolowski2018,Macquart2019}.
While the broad-band spectral index of $\alpha=-1.0$ \citep{Sokolowski2018} is flatter than the ASKAP measurement of $\alpha=-1.5$ \citep{Macquart2019}, the corresponding intrinsic slope ($\alpha_{\rm int}=-2.4$) is steeper than that of ASKAP measurement ($\alpha_{\rm int}=-2.0$).
This is because the absorption is stronger at lower frequencies, i.e., $\nu_{\rm obs}=0.74$ GHz for the MWA-ASKAP joint constraint and 1.297 GHz for the ASKAP measurement.

\section{Effects of scattering, Galactic coordinates, and free-free absorption}
\label{effects}
\begin{figure*}
	\includegraphics[width=2.0\columnwidth]{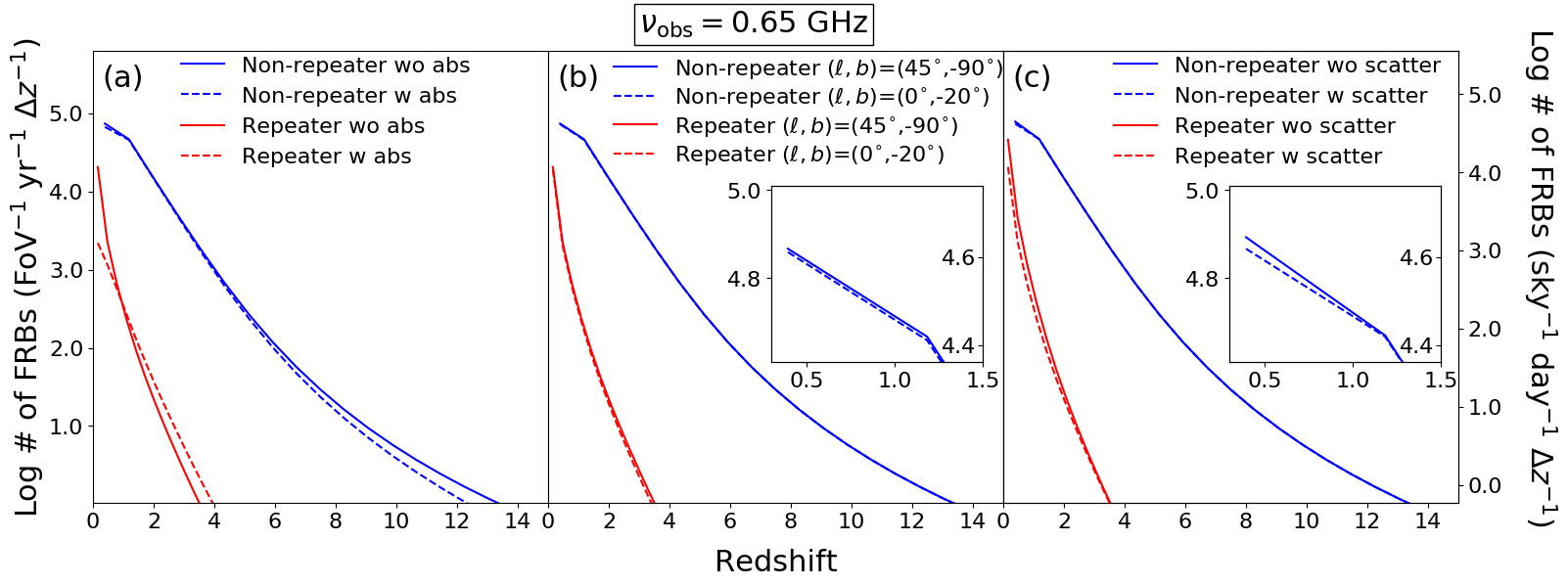}
    \caption{
    Differences between the predicted numbers of FRBs to be detected with the SKA with different assumptions.
    (Left) The dashed and solid lines represent predicted FRB detections with and without considering the free-free absorption, respectively.
    The spectral index is $\alpha=-1.5$ for the non-absorption cases and the intrinsic spectral index is $\alpha_{\rm int}=-2.0$ for the absorption-included cases. 
    The blue and red colours correspond to non-repeating FRBs and sources of repeating FRBs, respectively.
    We assume $(\ell,b)=(45^{\circ}.0,-90^{\circ}.0)$, $\nu_{\rm obs}=0.65$ GHz, integration over $\log L_{\nu}=25.0$-33.0 (erg Hz$^{-1}$) in Eq. \ref{N}, with scattering broadening, and the cosmic stellar mass-density (CSMD) evolution.
    (Middle) Same as left, except for the different assumptions on the Galactic coordinates.
    No absorption and $\alpha=-1.5$ are assumed.
    The inner box is the magnified knee of the non-repeating FRBs at $z\sim0.7$.
    (Right) Same as middle, except with and without scattering broadening.
    Both the dashed lines are calculated using weight factors of 0.7 and 0.3, as described in Section \ref{analysis}.
    }
    \label{figB1}
\end{figure*}

In Fig. \ref{figB1}, we demonstrate how free-free absorption, changing the Galactic coordinates, and scattering broadening, affect the predicted numbers of FRB sources to be detected with the SKA.
In the left panel, the non-absorption cases with $\alpha=-1.5$ are compared with the absorption cases with $\alpha_{\rm int}=-2.0$, following the characteristic spectral index of ASKAP FRBs \citep{Macquart2019}.
This is because the spectral index measured for ASKAP FRBs \citep{Macquart2019} corresponds to either $\alpha=-1.5$ without absorption or $\alpha_{\rm int}=-2.0$ with absorption.
While the absorption changes $L_{\nu}$, spectral shape, and luminosity functions of repeating FRBs, the differences in FRB detections are moderate compared to the different assumptions on the redshift evolution as shown in Figs. \ref{fig2}-\ref{fig3}.
In the middle panel, changing the Galactic coordinates slightly changes the FRB detections. 
This is because DM$_{\rm obs}$ is higher at lower Galactic latitudes, leading to the broadening of the FRB pulse via dispersion smearing and thus decreasing its signal-to-noise ratio.
The differences in FRB detection rates are less than 3\% for non-repeating FRBs and less than 20\% for repeating FRBs at 0.65 GHz. 
This effect is small because the intra-channel width of SKA is designed to be small enough \citep[$\Delta \nu_{\rm obs}=10$ kHz;][]{Torchinsky2016}.
In the right panel, the scattering broadening slightly changes the FRB detections.
The effect of scattering is even smaller at higher redshifts, since the observed frequency shifts to higher frequencies at the source frame whose scattering is smaller due to its $\nu_{\rm rest}^{-4}$ dependency \citep{Bhat2004}.
We caution that the assumptions on scattering broadening in this work could change when more data becomes available in the future.
Sources of repeating FRBs are relatively more affected by the different assumptions on Galactic coordinates and scattering than non-repeating FRBs.
This is because repeating FRBs are largely dominated by fainter populations, compared to non-repeating FRBs.
Such faint FRBs are more easily affected by the effects of dispersion smearing and scattering broadening.

\bsp	
\label{lastpage}
\end{document}